# How important is the acute phase in HIV epidemiology?

Brian G. Williams

South African Centre for Epidemiological Modelling and Analysis (SACEMA),
Stellenbosch, Western Cape, South Africa

Correspondence should be addressed to BW at BrianGerardWilliams@gmail.com

## Abstract

At present, the best hope for eliminating HIV transmission and bringing the epidemic of HIV to an end lies in the use of anti-retroviral therapy for prevention, a strategy referred to variously as *Test and Treat* (T&T), *Treatment as Prevention* (TasP) or *Treatment centred Prevention* (TcP). One of the key objections to the use of T&T to stop transmission concerns the role of the acute phase in HIV transmission. The acute phase of infection lasts for one to three months after HIV-seroconversion during which time the risk of transmission may be ten to twenty times higher, per sexual encounter, than it is during the chronic phase which lasts for the next ten years. Regular testing for HIV is more likely to miss people who are in the acute phase than in the chronic phase and it is essential to determine the extent to which this might compromise the impact of T&T on HIV-transmission.

Here we show that 1) provided the initial epidemic doubling time is about 1.0 to 1.5 years, as observed in South Africa, random testing with an average test interval of one year will still bring the epidemic close to elimination even if the acute phase lasts for 3 months during which time transmission is 26 times higher than in the chronic phase; 2) testing people regularly at yearly intervals is significantly more effective then testing them randomly; 3) testing people regularly at six monthly intervals and starting them on ART immediately, will almost certainly guarantee elimination.

In general it seems unlikely that elevated transmission during the acute phase is likely to change predictions of the impact of treatment on transmission significantly. Other factors, in particular age structure, the structure of sexual networks and variation in set-point viral load are likely to be more important and should be given priority in further analyses.

## Introduction

We can estimate the relative risk of infection during the acute and chronic stages of infection in two ways. We can estimate it indirectly if we have data on the viral load in the acute and chronic stages and data on the probability of a transmission event as a function of viral load.[1-6] We can estimate it directly if we have a sero-incident cohort with a short follow up time and if we can measure the number of infection events in each stage.[8] Both methods require a sero-incident cohort and repeated measurements of the viral load. Cohort data are always difficult to obtain and cohorts of discordant couples will become increasingly biased with time as those that are most likely to infect their partners do so and are removed from the cohort. Nevertheless, these are the data that we have.

## Indirect estimates

We first ask: how does viral load vary from initial infection through sero-conversion and the acute phase to the chronic phase and eventually to the final phase shortly before death? The most useful data in this regard, based on two sets of archived samples from HIV infected plasma donors, are from Fiebig *et al.*[7] Figure 1 shows the observed median values of viral load as a function of time since infection.[7] The fitted (green) line is

$$V = N\{l(t|\alpha_1,\delta_1) \times [\beta + (1-\beta)l(t|-\alpha_2,\delta_2)]\} \quad 1$$

where $V$ is the viral load, $t$ is the time since infection and

$$l(t|\alpha,\delta) = \frac{e^{\alpha(t-\delta)}}{1+e^{\alpha(t-\delta)}}. \quad 2$$

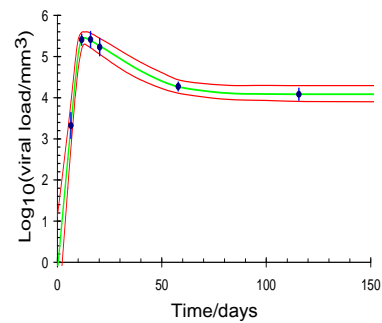

Figure 1. Median values of viral load as a function of time since infection.[7] Error bars are 95% confidence limits on median values. Green lines: maximum likelihood fit (see text); red lines 95% confidence bands. Data for individuals (not shown) vary by about ±1.5 logs.

In Equation 2, $N$ scales overall transmission up or down, the first logistic function $l$ in the curly brackets increases with time at a rate $\alpha_1$ reaching half the maximum value at time $\delta_1$, the second decreases with time at a rate $\alpha_2$ reaching half the maximum value at time $\delta_2$. $\beta$ is the asymptotic value to which the viral load converges during the chronic phase. In short, the viral load increases at a rate $\alpha_1$ to a peak value from which it converges downwards at a rate $\alpha_2$ to an asymptote at $\beta$. The median value of the $\log_{10}$(viral load) during the chronic phase is 4.09 ± 0.31 and at the peak of the acute phase is 5.45 ± 0.15. The acute phase lasts from day 8 to day 70 after sero-conversion so that the duration of the acute phase, $D_{AP}$ = 62 days, during which time the average value of the median viral load is 4.73 ± 0.02 for an average increase of 0.7 logs or a factor of $10^{0.7}$ = 5.0 (3.2–7.9) over the value during the chronic stage of infection.

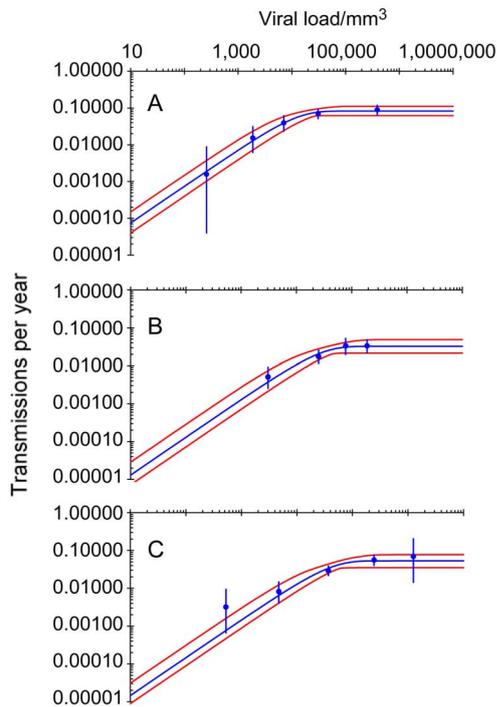

Figure 2. HIV transmission probability per year as a function of viral load. A: Attia et al.;[1] B: Donnell et al.;[3] C: Lingappa et al.[6] Significance levels for the fitted lines are A: 0.778; B: 0.376; C: 0.110. If we assume a power law relationship between transmission and viral load the significance levels for the best fit curves are A: 0.0006; B: 0.302; C: 0.309.

There are three sets of data on the risk of infection per unit time as a function of viral load[1,3,6] (Figure 2). The simplest model of the relationship between viral load and transmission assumes that the former increases linearly with the latter (but see also Appendix 1). However, the data suggest that transmission saturates at high viral load and we therefore assume a relationship of the following form:

$$T = \alpha\left(1 - e^{-\rho V}\right) \qquad 3$$

where $T$ is the probability of transmission per year. At low viral loads transmission increases linearly as $T = \alpha\rho V$ so that the probability of infection per virion per mm$^3$ is $\alpha\rho$, transmission saturates at $\alpha$ transmissions per year and we can define $V^*$, the viral load time at which transmission saturates, as the intercept of the initial linear increase with the asymptotic value so that

$$V^* = \frac{1}{\rho}. \qquad 4$$

The fits in Figure 2 give the values in Table 1. Allowing for the small amount of over-dispersion in the estimates we see that for low viral loads the probability of infection is $2.42 \times 10^{-6}$ ($1.14 \times 10^{-6}$–$5.13 \times 10^{-6}$) times the viral load per mm$^3$ and at high viral loads transmission saturates when the viral load is 4.37±0.29 logs.

We can now estimate the probability of transmission at the peak of the acute phase, when the $\log_{10}$ (median viral load) is 5.45, and in the chronic phase, when the $\log_{10}$ (median viral load), using each of the three data sets. The ratio of the pairs of estimates gives $RR$, the relative risk of infection, per unit time, in the acute and the chronic phase as $RR$ = 2.1 (1.1–3.9).

Table 1. The viral load at which transmission saturates; the probability of infection per virion per mm$^3$; and the relative risk ($RR$) of infection in the acute and chronic phases.

| Reference | $\log_{10}$(sat. viral load/mm$^3$) | (Prob. infection/ virion/mm$^3$/yr) ×10$^6$ | $RR$ acute v. chronic phase |
|---|---|---|---|
| Attia[1] | 4.05 ± 0.44 | 7.46 (4.00–14.9) | 1.4 (0.6–3.5) |
| Donnell[3] | 4.40 ± 0.39 | 1.30 (0.72–2.86) | 2.5 (0.7–8.6) |
| Lingappa[6] | 4.56 ± 0.36 | 1.46 (0.89–3.11) | 3.4 (1.0–11.3) |
| Average | 4.37 ± 0.29 | 2.42 (1.14–5.13) | 2.1 (1.1–3.9) |

## Direct estimates

The most widely cited direct estimates of the relative transmission in the acute and chronic stages of HIV-infection are based on the Uganda study of Wawer et al.[8] The most reliable data, in this regard, are those from the incidence cohort in which there were 10 transmission events among 23 couples who had 1221 coital acts in the first six months after seroconversion and 2 transmission events among the remaining 13 couples who had 1313 coital acts in months 6 to 15 after seroconversion. This gives a relative risk of transmission $RR$ = 3.4 (0.7–17.6). Wawer et al.[8] give an unadjusted estimate of the $RR$ of transmission, comparing the acute phase to the 'prevalent' cases, of 8.25 (3.37–20.22) but since the risk of transmission in 'prevalent' cases is close to half of the risk in incident cases 6 to 15 months after sero-conversion, their estimate of the $RR$, using those in the incidence cohort who were infected 6 to 15 months after sero-conversion, would be 4.1 (1.6–10.1). This is still significantly higher than the indirect estimate given above, especially since the Wawer et al.[8] estimate is averaged over six months while the estimate made here is averaged over two months. In order to favour the importance of the acute phase on the epidemic we will use the high estimate from Wawer et al.[8]

Now, consider a cohort of newly infected people. Viral load varies considerably among people infected with HIV. From data on the distribution of viral load in a cross-sectional study from Orange Farm, South Africa[12] (Bertran Auvert, personal communication) $\log_{10}$(viral load) measurements range from 2 to 6 corresponding to a range in the risk of transmission of 100 to 1000 times. Allowing for the fact that those with the highest viral load will infect their partners and die more quickly,[9] removing them from the pool of sero-discordant couples, the rate of transmissions will fall by about 50% after five years as shown in Figure 4 and observed by Wawer et al.[8] (See Appendix 2 for details.)

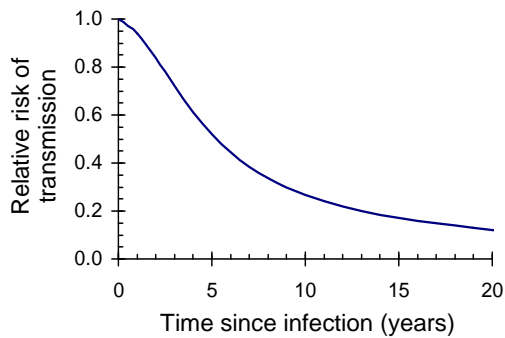

Figure 3. Variation of infectiousness as a function of time since infection in a cohort of people. No allowance is made for increased infectiousness during the acute phase but it is assumed that both infectiousness and mortality are highest in those with a high set-point viral load.

Hollingsworth et al.,[10] using the data provided by Wawer et al.,[8] use a modelling approach to determine the relative risk of transmission during the acute and chronic phases. Their estimate of the duration of the acute phase is $D_{AP}$ = 2.9 (1.2–6.0) months with $RR$ = 26.2 (12.5–53.5). At the limits we can assume that the short estimate of $D_{AP}$ corresponds to the high value of $RR$ while the long estimate of $D_{AP}$ corresponds to a low value of $RR$. Assuming that their comparison is with the sero-prevalent couples one might again reduce the relative risk by a factor of about 2. An inevitable consequence of these estimates is that $R_0$ must be close to 1 and indeed Hollingsworth et al.[10] give an estimate of 2.2 under random mixing. If this were the case HIV should be much less stable than it is observed to be and small improvements in prevention should have a substantial impact on the epidemic which is not seen to be the case. Again, for the purposes of this analysis, we will use the high estimate from Hollingsworth et al.[10]

The estimates of $D_{AP}$ and $RR$ that we will use in this analysis are given in Table 2.

Table 2. Estimates of the duration of the acute phase, $D_{AP}$, and the relative risk of transmission, $RR$, used in this analysis.

|  | $D_{AP}$ (mo.) | $RR$ acute/chronic phase |
|---|---|---|
| This study | 1 | 2.1 (1.1–3.9) |
| Wawer[8] | 6 | 8.2 (3.4–20.2) |
| Hollingsworth[10] | 2.9 (1.2–6.0) | 26.2 (12.5–53.5) |

## The impact of universal testing on transmission

We now wish to explore the consequences for the different estimates of the acute phase duration and the relative infectiousness of the acute phase on the impact of T&T on transmission. The key point is this: one of the few directly observed parameters concerning the epidemiology of HIV is the initial doubling time which, in South Africa, is 1.25 ± 0.25.[11] Since the acute phase lasts for considerably less time than the chronic phase, the greater the relative risk of transmission in the acute phase the smaller must be the value of $R_0$ to maintain the same initial doubling time. Indeed, if we know the initial doubling time (the growth rate $r$ in Equations 11 and 13 in Appendix 3) and we know the relative risk of infection in the acute phase and each of the four chronic phases ($\beta_i/\beta_0$ in Equations 7 to 9 in Appendix 3) and the duration of each of the four stages ($1/\rho_i$ in Equations 7 to 9 and 10) in Appendix 3, then we can determine the values of the each of the individual $\rho_i$ and hence the value of $R_0$ (Equation 10 in Appendix 3).

Values of $R_0$, as a function of the relative risk of transmission during the acute phase and the duration of the acute phase in months, are given in Figure 4A. Without ART the value of $R_0$ is 5.8 (brown rectangle). With $RR$ = 2.1 and $D_{AP}$ = 2 mo. (green ellipse) $R_0$ falls to 5.4. With $RR$ = 8.3 and $D_{AP}$ = 6 mo. (red ellipse) $R_0$ falls to 3.0. With $RR$ = 26 and $D_{AP}$ = 6 mo. (blue ellipse) $R_0$ falls to 2.3. As expected the higher the rate of transmission during the acute phase the lower the value of $R_0$.

The boundaries of the ellipses indicate the uncertainty in the point estimates which are considerable. Assuming, as noted above, that in the Hollingsworth et al.[10] study the high values of $D_{AP}$ correspond to low values $RR$, and vice versa, we slant the corresponding confidence ellipse at an appropriate angle. This also serves to show that if we let $D_{AP}$ = 6 mo. The Hollingsworth et al.[10] and the Wawer et al.[8] estimates are not significantly different.

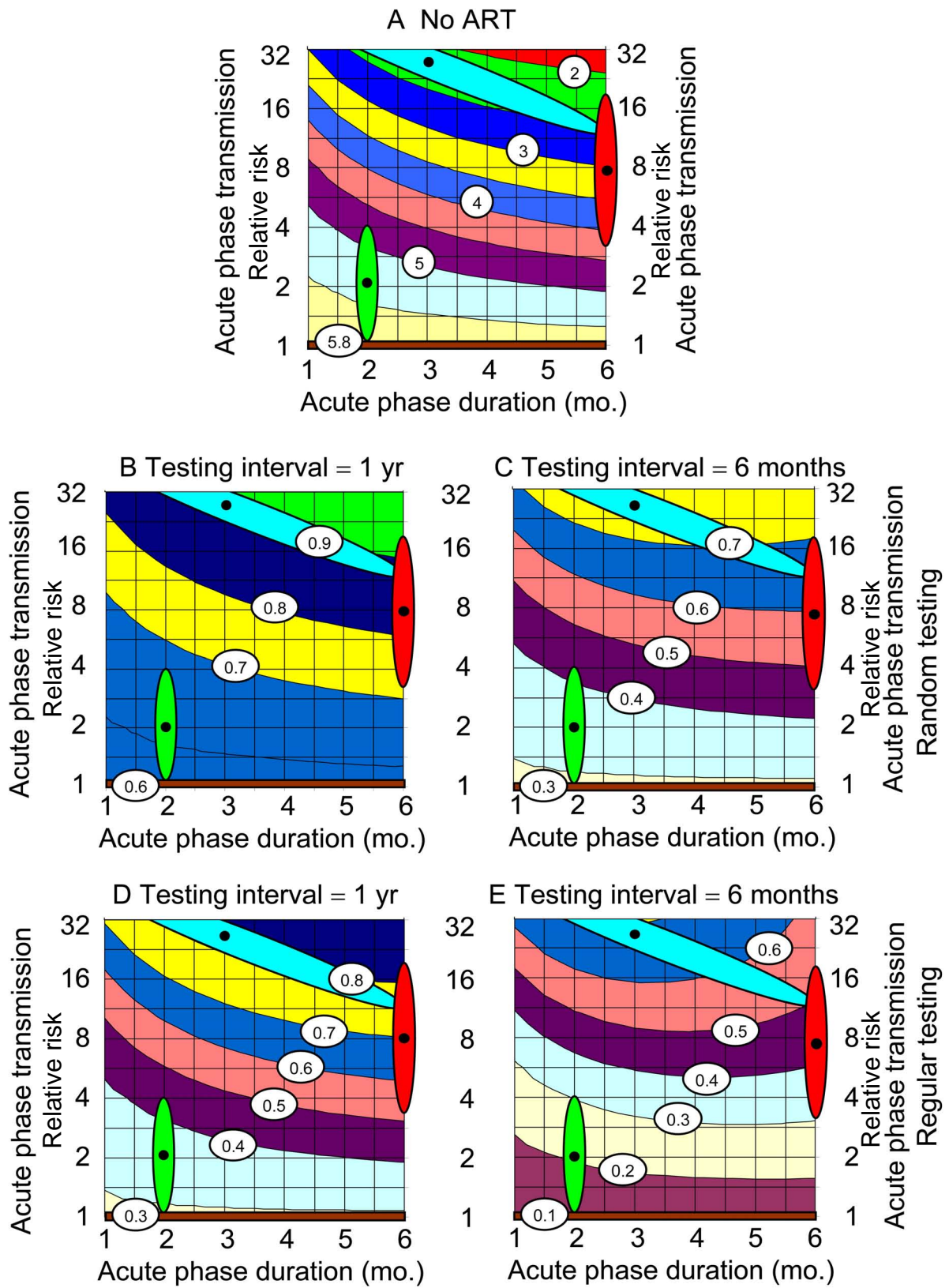

Figure 4. The value of $R_0$ for an epidemic with a doubling time of 15 months as a function of the duration of the acute phase and the relative transmission rate in the acute phase compared to the chronic phase. Colours correspond to different values of $R_0$ and the numbers in circles give the values on different contour lines. A without ART; B and C with random testing once a year or twice a year, on average; D and E with regular testing once a year or twice a year. Brown rectangle: $RR = 1$; green ellipse: $RR = 2.1$, $D_{AP} = 1$ mo.; ellipse $RR = 8.3$, $D_{AP} = 6$ mo.; blue ellipse: $RR = 26$, $D_{AP} = 3$ mo. The size of the ellipse indicates the uncertainty in the estimate.

Figure 4B shows what happens if people are tested randomly but once a year on average. With $RR = 1$, the value of $R_0$ falls to 0.58 (brown rectangle). With $RR = 2.1$ and $D_{AP} = 2$ mo. $R_0$ falls to 0.61. With $RR = 8.3$ and $D_{AP} = 6$ mo. $R_0$ falls to 0.84 and with $RR = 26$ and $D_{AP} = 3$ mo. $R_0$ falls to 0.90. In all three cases $R_0$ still falls below 1 although with the two higher estimates of $RR$ it is only 10% to 20% below 1 which allows little margin of error.

Figure 4C shows what happens if the average testing interval is reduced to six months. With $RR = 1$, $R_0$ falls to 0.29 (brown rectangle). With $RR = 2.1$ and $D_{AP} = 2$ mo. $R_0$ falls to 0.34. With $RR = 8.3$ and $D_{AP} = 6$ mo. $R_0$ falls to 0.61 and with $RR = 26$ and $D_{AP} = 3$ mo. $R_0$ falls to 0.75. Even in the worst case ($RR = 26$, $D_{AP} = 3$ mo.) $R_0$ is significantly less than 1.

Figure 4D shows what happens if people are tested regularly once a year. With $RR = 1$, the value of $R_0$ again falls to 0.29 (brown rectangle). With $RR = 2.1$ and $D_{AP} = 2$ mo. $R_0$ falls to 0.38. With $RR = 8.3$ and $D_{AP} = 6$ mo. $R_0$ falls to 0.70 and with $RR = 26$ and $D_{AP} = 3$ mo. $R_0$ falls to 0.82. Again, even in the worst case ($RR = 26$, $D_{AP} = 3$ mo.) $R_0$ is significantly less than 1.

Figure 4E shows what happens if people are tested regularly twice a year. With $RR = 1$, $R_0$ falls to 0.14 (brown rectangle). With $RR = 2.1$ and $D_{AP} = 2$ mo. $R_0$ falls to 0.21. With $RR = 8.3$ and $D_{AP} = 6$ mo. $R_0$ falls to 0.45 and with $RR = 26$ and $D_{AP} = 3$ mo. $R_0$ falls to 0.68. Even in the worst case ($RR = 26$, $D_{AP} = 3$ mo.) $R_0$ is again significantly less than 1.

**Conclusion**

We have three estimates of $RR$, the relative risk of infection, and $D_{AP}$, the duration of the acute phase ranging from 2.1 over 2 months to 26.2 over 3 months giving values of $R_0$ ranging from 5.8 to 2.3. The high estimates for $RR$ may well be over-estimates. They are both based on the data from Rakai[8] and discordant couple studies in which one partner is 'sero-prevalent' will select against those couples who are most infectious and therefore no longer sero-discordant. Furthermore, the high values of $RR$ with long values of $D_{AP}$ imply values of $R_0 \approx 2$ which seems unlikely; if this were the case HIV should be relatively easy to eliminate through minor changes in behaviour and the epidemic should be much less stable.

However, it is clear from this analysis that even if we adopt the most pessimistic view and assume the relative risk of infection is 26 times higher during an acute phase that lasts for 3 months annual testing and immediate treatment has the potential to $R_0$ to less than 1 and with any further contribution to prevention will probably guarantee elimination in the long term. Testing people regularly, on an annual basis, is considerably more effective than random testing because under random testing some people will be tested very frequently, which is not necessary, while others will be tested very infrequently which is not ideal. With regular testing even the most pessimistic view reduces $R_0$ to 0.82 and will probably lead to elimination. As expected, testing people twice a year reduces $R_0$ even further and under all assumptions about the acute phase would guarantee elimination.

However, this analysis raises a possibility that may be even more important than considerations of high rates of infection during the acute phase. We know that the set point values of the viral load vary by several orders of magnitude. In the Fiebig[7] study the $\log_{10}$(viral load) varies from about 2.6 to 5.6 in the chronic stage.

The data in Figure 4 suggest that variation in the average viral load in a cohort of people, as a function of time since infection, is likely to have a greater influence on the model predictions than any difference between acute and chronic phase transmission. However, there are two reasons why this is more difficult to allow for. First of all the result shown in Figure 4 assumes a relationship between survival and set-point viral load based on only one small study and better data are needed if this is to be made the basis for modelling the epidemic. Secondly, including this variation would probably need a model that includes the distribution of set-point viral loads explicitly and the structure of the models that are currently used does not allow for this. If this were to be explored further, the first priority would be to consider models in which the variation in the set-point viral load is included explicitly and comparisons made with a model in which this variation is set to zero.

We conclude, therefore, that increased transmission during the acute phase is unlikely to change the model predictions significantly for several reasons:
1. The acute phase duration is more likely to be of the order of one or two months or about 1% to 2% of the total disease duration.
2. If transmission during the acute phase is sufficiently high for transmission during this short time to be important, then $R_0$ must be correspondingly low and the reduction in $R_0$ needed for elimination is correspondingly less.
3. There is strong evidence that transmission saturates above a viral load of about 4 to 5 logs mitigating the impact of even very high viral loads during the acute phase.
4. A more important limitation of the current model structures, is that variation in the set-point viral load is not included and this should be explored further.

5. The short duration of the acute phase means that it can only ever make a significant contribution to transmission if the rate of partner changes is much higher than is generally observed to be the case.

## Appendix 1. Fitting transmission as a power-law function of viral load

Several authors[13] have fitted the relationship between transmission and viral load to a power law function with the transmission increasing as the viral load to the power of about 0.3. This implies that at all viral loads transmission increases more slowly then linearly so that there is a degree of saturation in transmission as viral load increases. However, there is no obvious biological basis for this although from a statistical point of view it is not possible to choose between this power law model and the linear increase to an asymptote suggested here.

## Appendix 2. Reduction in transmission

Let the relative risk of infection vary with time since infection as $RR(t)$ Then under random testing at a rate $\rho$ year, the reduction in the overall transmission will be

$$R = \frac{\int_0^\infty e^{-\rho t} RR(t)\,dt}{\int_0^\infty RR(t)\,dt} \qquad 5$$

while under regular testing at an interval of $\tau$ years the reduction in overall transmission will be

$$R = \frac{\int_0^\tau \left(1 - \frac{t}{\tau}\right) RR(t)}{\int_0^\infty RR(t)} \qquad 6$$

Since we have estimates of the relative risk of transmission, given that a person is alive, for different stages of infection, we approximate $RR(t)$ with an appropriate step function.

## Appendix 3. $R_0$ and the growth rate for an $n$-stage model

We want to introduce an acute phase but also keep four chronic stages in order to ensure that the survival distribution approximates the observed Weibull distribution with a shape parameter of abut 2 reasonably well. The equations for this model are (keeping the total population constant)

$$\dot{I}_0 = -I_0 \sum_{i=1}^n \beta_i I_i + \rho_n I_n \qquad 7$$

$$\dot{I}_1 = I_0 \sum_{i=1}^n \beta_i I_i - \rho_1 I_1 \qquad 8$$

$$\dot{I}_i = \rho_{i-1} I_{i-1} - \rho_i I_i, \quad i = 2 \text{ to } n \qquad 9$$

where $I_0$ refers to the proportion of people that are susceptible people and $I_i$ to the proportion in each successive stage of infection. The infectiousness of each stage is determined by $\beta_i$ and the mean duration of each stage is $1/\rho_i$. From Equations 7 to 9 it follows that

$$R_0 = \sum_{i=1}^n \frac{\beta_i}{\rho_i} \qquad 10$$

During the initial exponential growth of the epidemic at a rate $r$ we have $I_0 \approx 1$ and

$$\frac{\dot{I}_i}{I_i} = r, \quad i = 0 \text{ to } n \qquad 11$$

from which it follows that

$$\sum_{i=1}^n \beta_i r_i = 1 \qquad 12$$

with

$$r_i = \prod_{j=0}^{i-1} \frac{\rho_j}{r + \rho_j}. \qquad 13$$

We can now determine the relationship between the relative risk of infection in each stage and $R_0$ while constraining the overall growth rate, $r$, as follows. We first set the duration of each stage $1/\rho_i$ and the relative infectiousness of each stage $\beta_i/\beta_0$, for $i = 1$ to $n$, and then vary $\beta_0$ to get the required value of the initial growth rate $r$. Using Equation 10 we then calculate $R_0$ directly.